\documentclass[aps,prb,twocolumn]{revtex4}
\usepackage{graphicx}
\usepackage{amsmath}
\usepackage{amsfonts}
\usepackage{times}
\usepackage{bm}

\begin{document}

\title{The single-mode description of the integer quantum Hall state 
with dipole-dipole interaction}

\author{Rui-Zhi Qiu$^{1,2}$, Zi-Xiang Hu$^{3}$, and Xin Wan$^{1}$}
\affiliation{$^1$Zhejiang Institute of Modern Physics, Zhejiang University,
Hangzhou 310027, P.R. China}
\affiliation{$^2$Science and Technology on Surface Physics and Chemistry 
Laboratory, P.O. Box 718-35, Mianyang 621907, Sichuan, P.R. China}
\affiliation{$^3$Department of Physics, Chongqing University,
Chongqing 400044, P.R.China}

\date{\today}

\begin{abstract}

A topological phase can often be represented by a corresponding
wavefunction (exact eigenstate of a model Hamiltonian) that has a
higher underlying symmetry than necessary. When the symmetry is
explicitly broken in the Hamiltonian, the model wavefunction fails to
account for the change due to the lack of a variational parameter.
Here we exemplify the case by an integer quantum Hall system with
anisotropic interaction. We show that the single-mode approximation
can introduce a variational parameter for a better description of the
ground state, which is consistent with the recently proposed geometric
description of the quantum Hall phases.

\end{abstract}

\maketitle

\section{Introduction}

The integer quantum Hall (IQH) effect is now regarded as a simple
class of topological insulators with broken time-reversal symmetry. A
system in such a phase has invariant properties of its ground state
wavefunction under smooth deformations of the Hamiltonian. This, of
course, does not mean that the ground state wavefunction itself is
invariant. For convenience, we often choose to study the system with
continuous symmetries in translation and/or rotation in which the
ground state wavefunction is relatively simple, although these
symmetries are unnecessary for the study of the topological
properties. In fact, such a choice can sometimes hide important
physics.

Consider an IQH system in the disk geometry with rotational
symmetry. Electrons in a polarized lowest Landau level (LLL) fill up
ring-shaped orbitals $\phi_m(z) = z^m e^{-|z|^2/4}$, where $z = x +
iy$ is the complex coordinate on the plane. The trivial many-body
ground state is the Slater determinant of the orbitals with the lowest
possible $m$
\begin{equation}
\label{eq:iqhe}
\Phi_0(z_1,...,z_N) = \left [ \prod_{i < j} (z_i - z_j) \right ]
e^{-\sum_i |z_i|^2/4},
\end{equation}
which is the well known Vandermonde determinant multiplied by the
ubiquitous Gaussian factor. The wavefunction, which vanishes when two
electrons coincide with each other, manifests the Pauli exclusion
principle. The trivial many-body state is known to be the ground state
even in the presence of interaction.

This prompts some interesting questions. Can effects of interaction be
explicitly manifested in the robust IQH state? If yes, how do they
modify the ground state wavefunction variationally? These questions
can be likewise asked for the fractional quantum Hall (FQH) states,
which have been explored in recent experiments using tilted magnetic
fields.~\cite{xia11,kamburov13} In fact, such attempts have been made
recently in the context of the geometric description of FQH
states,~\cite{haldane11,qiu12} which can be applied straightforwardly
to the IQH counterpart.

To construct a variational wavefunction for correlated systems from
its noninteracting counterpart is a long-standing
problem.~\cite{fuldebook} Suppose we decompose a many-body Hamiltonian
to be $H = H_0 + H_1$. We assume that we know the ground state $\vert
\Phi_0 \rangle$ of $H_0$, which describes, for the present discussion,
the noninteracting part of the total Hamiltonian. $H_1$ describes the
interparticle interaction. How can the ground state wavefunction be
variationally constructed? Within the single-mode approximation, which
replaces the effect of $H_1$ by a single, mean excitation energy
$\omega_0$, we can write down an ansatz wavefunction $\vert \Psi_0
\rangle = e^{-H_1 / \omega_0} \vert \Phi_0 \rangle$ for the ground
state of $H$.~\cite{fuldebook} Here, $\omega_0$ can be tuned as a
variational parameter to minimize the ground state energy. A
well-known example is the Gutzwiller's
wavefunction,~\cite{gutzwiller63} which accounts for the electron
correlation in the on-site Hubbard model. An improved independent-mode
approximation can lead to Jastrow's ansatz for a trial ground-state
wavefunction
\begin{equation}
\label{eq:jastrow}
\Psi_0 \left (\mathbf{r}_1,...,\mathbf{r}_N \right ) = e^{
\sum_{ij} f \left (\mathbf{r}_i - \mathbf{r}_j \right ) }
\Phi_0 \left (\mathbf{r}_1,...,\mathbf{r}_N \right ),
\end{equation}
where the Jastrow function $f(\mathbf{r} - \mathbf{r'})$ is determined
by energy minimization.~\cite{fuldebook}  

According to the single- and independent-mode approximations, the
Jastrow function for the IQH state simply vanishes despite the
interparticle interaction. This opinion, of course, is
counterintuitive hence oversimplified, especially in the case with
anisotropic interaction or mass tensor, as was already discussed by
Haldane~\cite{haldane11} in the FQH context recently. In this article
we take an empirical approach to decipher the IQH ground-state
wavefunction from numerical diagonalization of a microscopic system
with dipole-dipole interaction. We find that in the IQH regime the
realistic ground state can be written as the isotropic IQH ground
state multiplied by a sum of a few polynomials, which can be cast in
an exponential form much like the Jastrow factor in
Eq.~(\ref{eq:jastrow}). The Jastrow factor can be understood as
arising, in the single-mode approximation, from an anisotropic
quadrupolar interaction. Up to a center-of-mass contribution
suppressed by the one-body potential, we show that the wavefunction is
consistent with the Bogoliubov transformation of the guiding center
coordinates in the isotropic wavefunction.  The paper is organized as
follows. We introduce our model on quantum Hall systems with
dipole-dipole interaction in Sec.~\ref{sec:model}. In
Sec.~\ref{sec:wavefunction} we analyze the structure of the ground
state wavefuntion and discuss the regime where it can be understood as
a single-mode approximation. We discuss the connection of our findings
to recent works on the geometric description on quantum Hall states in
Sec.~\ref{sec:discussion}. We summarize the paper in
Sec.~\ref{sec:summary} and leave some technical details of our
calculation in Appendix~\ref{app:proof}.

\section{Model}
\label{sec:model}

Recent experimental and theoretical development in both dipolar atoms
and polar molecules~\cite{baranov12} promises us the potential of
realizing the FQH effect in systems with tunable and anisotropic
interaction.\cite{cooper05,baranov05,grass11} In this section we
review our model for dipolar atomic systems in the quantum Hall regime
and identify the phase of interest in this study.

As proposed in Ref.~\onlinecite{qiu11}, the fast rotating
quasi-two-dimensional gas of polarized fermionic dipoles serves as an
ideal arena for the present study.  The Larmor theorem states that the
fast rotation is equivalent to a high magnetic field for the fermions.
Since the p-wave interaction for the polarized fermions is typically
small unless in the resonance regime, the only significant interaction
in the system is the dipole-dipole interaction.  Here without loss of
generality, we assumed that the dipole moments are polarized in the
$x$-$z$ plane of the rotating frame and the motion of all dipoles
along $z$-axis is frozen to the ground state of the axial harmonic
oscillator, $\phi_z(\zeta)=\pi^{-1/4}q^{-1/2}e^{-\zeta^2/(2q^2)}$.
The quasi-two-dimensional dipole-dipole interaction is described by
\begin{eqnarray}
{\cal V}_\theta^{(2D)}(x,y) &=& c_d V_\theta^{(2D)}(x,y), \mbox{ where}\nonumber\\
V_\theta^{(2D)}(x,y) &=& \frac{1}{(2\pi q^2)^{1/2}}
\int d\zeta e^{-\zeta^2/(2q^2)}\nonumber\\
&\times&\frac{x^2+y^2+\zeta^2-3(\zeta\cos \theta+x\sin\theta)^2}
{(x^2+y^2+\zeta^2)^{5/2}},\nonumber
\end{eqnarray}
$c_d$ is the interaction strength and $\theta$ is the polar angle of
the dipole moment.

Therefore, within the LLL formalism, we set up a quantum Hall system
with anisotropic dipole-dipole interaction as described by the
following Hamiltonian
\begin{eqnarray}
H=\alpha L^z + \frac{1}{2}\sum_{m_1,m_2,m_3,m_4}V_{1234}(\theta)
f^\dag_{m_1}f^\dag_{m_2}f_{m_4}f_{m_3},\label{eq:hamiltonian}
\end{eqnarray}
where $f_m^\dag$ creates a fermion in the state $\phi_m$ and
$L^z=\sum_m mf^\dag_m f_m$ is the $z-$component of the total angular
momentum.  The interaction matrix elements $V_{1234}(\theta)$ are
given by
\begin{eqnarray}
\int &d{\bm r}d{\bm r}'& 
\psi_{m_1}({\bm r})
\psi_{m_2}({\bm r}')
V^{(2D)}_\theta({\bm r}-{\bm r}')
\psi_{m_3}({\bm r})
\psi_{m_4}({\bm r}')\nonumber\\
&=&\cos^2\theta\, V^{z}_{1234}+ 
\sin^2\theta \, V^{x}_{1234},\label{eq:v1234}
\end{eqnarray}
where 
\begin{eqnarray}
V^z_{1234}&=&\delta_{m_1+m_2,m_3+m_4}\left[ -4{\cal A}_{1234}{\cal K}_{1234} 
\right.\nonumber\\
&&\left.+\frac{1}{3q}\sqrt{\frac{2}{\pi}}
\frac{(m_1+m_2)!}{2^{m_1+m_2}\sqrt{m_1!m_2!m_3!m_4!}}\right],\nonumber\\
V^x_{1234}&=&-\textstyle{\frac{1}{2}}\,V^z_{1234}+{\cal A}_{1234}
{\cal K}_{1234}\delta_{m_1+m_2,m_3+m_4\pm2},\nonumber
\end{eqnarray}
and
\begin{eqnarray}
{\cal A}_{1234}&=&\frac{1}{4\sqrt{2\pi}q}
\frac{i^{|m_3-m_1|-|m_4-m_2|}}{2^{(|m_3-m_1|+|m_4-m_2|)/2}}
\sqrt{\frac{[m_{13}^<]![m_{24}^<]!}{[m_{13}^>]![m_{24}^>]!}},
\nonumber\\
{\cal K}_{1234}&=&\int dt t^{(|m_3-m_1|+|m_4-m_2|)/2}e^{-t}
L_{m^<_{13}}^{|m_3-m_1|}\left(\frac{t}{2}\right)\nonumber\\
&\times&L_{m^<_{24}}^{|m_4-m_2|}\left(\frac{t}{2}\right){\cal E}\left(
q\sqrt{\frac{t}{2}}\,\right).\nonumber
\end{eqnarray}
Here $m_{ij}^<={\rm min}(m_i,m_j)$, $m_{ij}^>={\rm max}(m_i,m_j)$,
$L^n_m(\cdot)$ is the associated Laguerre polynomial, and ${\cal
  E}(x)=\sqrt{\pi}x e^{x^2}{\rm erfc}(x)$ with ${\rm erfc}(\cdot)$
being the complementary error function.  From Eq.~(\ref{eq:v1234}), it
is clear that the matrix elements $V_{1234}(\theta=0)$ are nonzero for
$m_1 + m_2-m_3-m_4 = 0$ and $V_{1234}(\theta\neq0)$ are nonzero when
$m_1 + m_2 -m_3-m_4 = 0$ or $\pm2$. 

The details of the potential physical realization of the Hamiltonian
were explained in Ref.~\onlinecite{qiu11}.  The isotropic confining
potential strength $\alpha$ can be tuned by the rotation frequency and
the tilt angle $\theta$ by applied electric and magnetic field.  We
explore, by exact diagonalization, the $N$-particle ground state and
compute its mean total angular momentum (in units of $\hbar$)
\begin{eqnarray}
\overline M\equiv\left\langle\Psi^{(N)}(\alpha,\theta)\right|L^z
\left| \Psi^{(N)}(\alpha,\theta)\right\rangle
\end{eqnarray}
where $\Psi^{(N)}(\alpha,\theta)$ denotes the ground state wave
function for given parameters $\alpha$ and $\theta$.

\begin{figure}[tbp]
\centering
\includegraphics[width=3.2in]{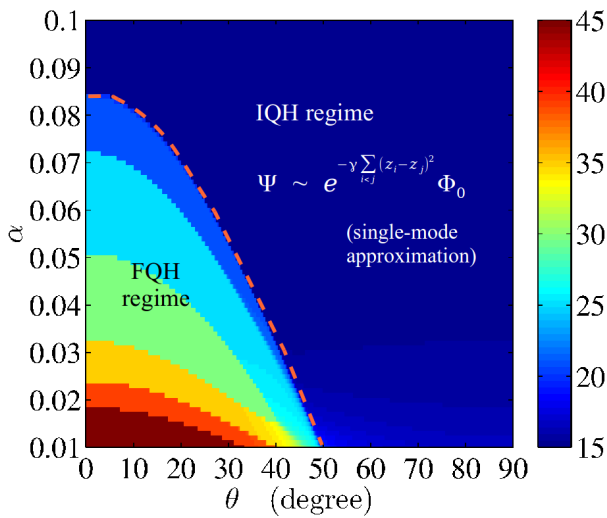}
\caption{(Color online) The mean total angular momentum of the ground
  state $\overline M = \langle\Psi^{(N)}(\alpha,\theta) \vert L^z
  \vert \Psi^{(N)}(\alpha,\theta) \rangle$ (in units of $\hbar$) in a
  system with $N=6$ particles for various confining potential strength
  $\alpha$ and tilt angle $\theta$.  The blue region to the top and
  right of the orange dashed line represents the anisotropic IQH
  state, while the rest is loosely referred to as the FQH regime, with
  the brown region in the lower left corner representing the $\nu =
  1/3$ Laughlin state. }
\label{phase}
\end{figure}

As illustrated in Fig.~\ref{phase}, with isotropic dipole-dipole
interaction in the plane of motion, i.e. for $\theta=0^\circ$ , the
total angular moment is a good quantum number and can be matched to
that of the IQH state $M_0 = N (N-1)/2$ for $\alpha > 0.085$ and $M_L
= 3N (N-1)/2$ the $\nu = 1/3$ Laughlin state for $\alpha<0.018$.  Note
that we only show $\alpha > 0.01$ to avoid the unwanted complication
due to weak confinement and finite system size.  As $\theta$
increases, the rotational symmetry is explicitly broken and $\overline
M$ increases from $M_0$ to reflect the anisotropy of the interaction.
We identify the extended blue region to the up and right of the curved
orange dashed line in Fig.~\ref{phase} to be the anisotropic IQH
phase with ${\overline M} = M_0 + O(1)$, which has the maximum density
in the interior of the oval-shaped droplet.\cite{qiu11}

The introduction of interaction anisotropy breaks the rotational
symmetry otherwise present in the disk geometry. This means that we
cannot diagonalize the Hamiltonian in the momentum subspace, hence the
calculations are restricted to rather small systems. But we have
checked out results for $N = 5$-7 particles and we believe the results
presented in the following sections are robust in even larger systems.

\section{Decipher the anisotropic ground-state wavefunctions}
\label{sec:wavefunction} 

The isotropic IQH ground state (in a harmonic trap) is the maximum
density droplet of electrons in the LLL. One expects that in the
presence of small anisotropic interaction the IQH state can only be
perturbed at the edge. Hence the anisotropic IQH state is expected to
be the superposition of the isotropic IQH ground state and its edge
states. In the following we will first describe the complete basis of
edge states in the isotropic case, which will be used later to
understand the ground-state wavefunctions in the anisotropic case.

\subsection{A complete basis in terms of edge states in the isotropic case}

Quantum Hall systems have gapless edge states. In the isotropic $\nu =
1$ integer quantum Hall case the edge states are simply generated by a
branch of chiral bosonic modes, which describe the deform of the
incompressible quantum Hall droplet at the edge. In the simpler
noninteracting case they are generated by electron-hole pair
excitations. In terms of analytical wavefunctions they can be written
down as the linear combinations of
\begin{equation}
  \left ( e_1^{n_1} e_2^{n_2} \cdots \right ) \Phi_{0} = 
\left (\prod_{m>0} e_m^{n_m} \right )
  \Phi_{0}
\end{equation}
with
\begin{eqnarray}
  e_0 & = & 1 \\
  e_1 & = & \sum_i z_i \\
  e_2 & = & \sum_{i < j} z_i z_j \\
  e_3 & = & \sum_{i < j < k} z_i z_j z_k \\
  & \vdots &  \nonumber
\end{eqnarray}
This is a complete basis for antisymmetric electron wavefunctions,
which vanish when $z_i \rightarrow z_j$. The convention is when we
refer to edge states we focus on those with excitation momentum
$\Delta M = O (1)$, as oppose to bulk excitations with $\Delta M = O
(N)$.

To count the edge states, we can define a generation function
\begin{equation}
G(q) = \sum_{\Delta M} q^{\Delta M} \cdot N_{\Delta M}
\end{equation}
where $N_{\Delta M}$ is the number of ways to partition $\Delta M$, i.e.,
sets of $n_m$ such that
\begin{equation}
  \Delta M = \sum^{\infty}_{m = 0} mn_m
\end{equation}
It is easy to see that the generation function is
\begin{equation}
G(q) = \prod_{m = 1}^{\infty} \frac{1}{1 - q^m}.
\end{equation}
The number of edge states for $\Delta M = 0$, 1, 2, 3, 4, ... are
simply 1, 1, 2, 3, 5, ....

One should notice that the basis we introduced above is not
orthogonal. For later numerical analysis it is convenient to introduce
the Schur polynomial as an alternative basis. We start with the
integer partition, $N-1$, $N-2$, ..., 0, of the ground-state total
angular momentum $M_0 = N(N-1)/2$. The ground-state wavefunction
$\Phi_0$ is simply the Slater determinant
\begin{eqnarray}
{\mathfrak sl}_{\{N-1, N-2, ..., 0\}} (\{z_i\}) &=& \left \vert 
\begin{array}{cccc}
z_1^{N-1} & z_2^{N-1} & \cdots & z_N^{N-1} \\
z_1^{N-2} & z_2^{N-2} & \cdots & z_N^{N-2} \\
\vdots & \vdots & \ddots & \vdots \\
z_1^{0} & z_2^{0} & \cdots & z_N^{0} 
\end{array}
\right \vert \nonumber \\
& = & \prod_{i < j} (z_i - z_j), \nonumber
\end{eqnarray}
multiplied by the ubiquitous Gaussian factor.  Naturally, edge
excitations are Slater determinants with corresponding partitions of
$M = M_0 + \Delta M$, and/or their linear combinations. Since
different Slater determinants are different occupations of electron
orbitals and hence orthogonal to each other, it is convenient to
represent the edge states in terms of Schur polynomials (multiplied by
the isotropic ground-state wavefunction).  For any integer partition
$\Delta M = d_{1} + d_{2} + \cdots + d_N$, where $d_{1} \ge
d_2 \ge \cdots \ge d_N$, the Schur polynomial is defined as
\begin{equation}
S_{\{d_i\}} (\{z_i\}) = \frac{{\mathfrak sl}_{\{N-1+d_1, N-2+d_2, ...,
    0+d_N\}} (\{z_i\})} {{\mathfrak sl}_{\{N-1, N-2, ..., 0\}}
  (\{z_i\})}.
\end{equation}
For examples, 
\begin{eqnarray}
S_{1}(\{z_i\})&=& \sum_{i}z_i = e_1, \\
S_{2}(\{z_i\})&=& \left ( \sum_{i}z_i \right )^2 - \sum_{i<j}z_i z_j =
e_1^2 - e_2,\\ 
S_{1,1}(\{z_i\})&=&\sum_{i<j}z_i z_j = e_2.
\end{eqnarray}
where we neglect the trailing zeros in the partition $\{d_i\}$. This
Schur polynomial basis is found to be most convenient later when we
deal with wavefunction overlap and normalization.

\subsection{Wavefunctions for the edge states in the isotropic case 
with dipole-dipole interaction ($\theta = 0$)}

\begin{figure}[tbp]
\centering
\includegraphics[width=3.2in]{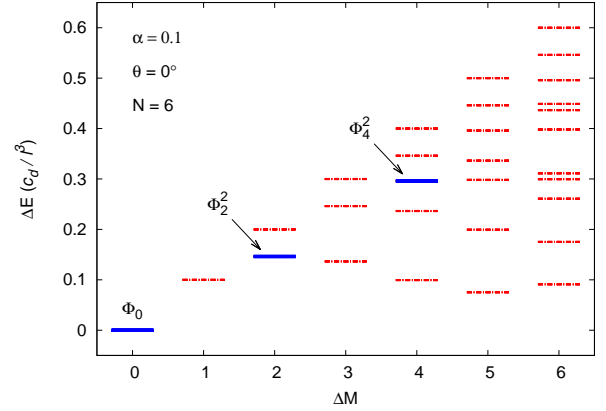}
\caption{(Color online) Low-energy excitations for the isotropic IQH
  state with $\alpha = 0.1$ and $\theta = 0^{\circ}$.  The blue thick levels
  can be identified as $\Phi_{0}$ [Eq.~(\ref{eq:iqhe})],
  $\Phi_{2}^{2}$ [Eq.~(\ref{eq:phi22})], and $\Phi_{4}^{2}$
  [Eq.~(\ref{eq:phi42})]. Note that their excitation energies $\Delta
  E$ scale with their momenta $\Delta M$. These states are the main
  components of the anisotropic IQH ground state for $\theta \neq
  0^{\circ}$. }
\label{IQHedge}
\end{figure}

Now we apply the Schur polynomial basis to analyze the isotropic
(i.e. $\theta = 0$) IQH ground state and its edge states. For
illustration we plot the low-energy spectrum in Fig.~\ref{IQHedge} for
$N = 6$ particles with isotropic dipole-dipole interaction and $\alpha
= 0.1$. The state at $\Delta M = 0$ is the isotropic ground state
$\Phi_0$, while the rest are the complete edge states up to $\Delta M
= 6$, as the countings in respective momentum sectors are 1, 1, 2, 3,
5, 7, 11 as expected. Bulk excitations are gapped by the Landau level
(LL) spacing, assumed to be infinity in the single LL projection.  For
$\Delta M = 1$ there is only one state and obviously
\begin{equation}
\label{eq:phi11}
\Phi_{1}^{1} = \sum_{i}z_i \prod_{i<j}(z_i-z_j) e^{-\sum_i|z_i|^2/4} =
S_{1} \Phi_0.
\end{equation}

For $\Delta M = 2$ we have two states, which should be linear
combinations of $S_2 \Phi_0$ and $S_{1,1} \Phi_0$. One can see that
the combinations
\begin{eqnarray}
\label{eq:phi21}
\Phi_{2}^{1}&=& \left [\sum_{i}z_i 
\right ]^2\prod_{i<j}(z_i-z_j) e^{-\sum_i|z_i|^2/4} \nonumber \\
&=& 
\left [S_2 + S_{1,1} \right ] \Phi_0, \\ 
\label{eq:phi22}
\Phi_{2}^{2}&=& \left [\sum_{i<j}(z_i-z_j)^2
  \right ]\prod_{i<j}(z_i-z_j) e^{-\sum_i|z_i|^2/4} \nonumber \\
&=& 
\left [(N-1)S_2 - (N+1)S_{1,1} \right ] \Phi_0
\end{eqnarray}
are the suitable choice, because the latter reduces the interaction
energy by allowing higher-degree zeros when any two particles approach
each other. One can thus identify the two states in the low-energy
spectrum; in particular, the state $\Phi_{2}^{2}$ is labeled in
Fig.~\ref{IQHedge} for later purposes.

One can extend the analysis to higher momentum, which we will not
continue here. But we should also highlight, in Fig.~\ref{IQHedge},
another edge state
\begin{eqnarray}
\label{eq:phi42} 
\Phi_{4}^2 &=& \left [\sum_{i<j}(z_i-z_j)^2
  \right ]^2 \prod_{i<j}(z_i-z_j) e^{-\sum_i|z_i|^2/4} \nonumber \\
&=& \left [ (N-1)^2S_{4}-(N-1)(N+3)S_{3,1} \right . \nonumber \\
& & +2(N^2+1)S_{2,2}
-(N-3)(N+1) S_{2,1,1} \nonumber \\
& & \left . +(N+1)^2S_{1,1,1,1} \right ] \Phi_0,
\end{eqnarray}
whose excitation energy doubles that of $\Phi_{2}^{2}$, again for
later purposes. The other four states in the same momentum sector are
other orthogonal linear combinations of $S_{4}\Phi_0$,
$S_{3,1}\Phi_0$, $S_{2,2}\Phi_0$, $S_{2,1,1}\Phi_0$, and
$S_{1,1,1,1}\Phi_0$. The highest level, e.g., can be identified as
$(\sum_i z_i)^4 \Phi_0$.

These edge state model wavefunctions are known in earlier
literature. For example, $\Phi_{2}^{2}$, as indicated in the low-lying
energy levels for the corresponding isotropic interaction in
Fig.~\ref{IQHedge}, is known to be the eigenstate of the projected
isotropic interaction in the lowest Landau level.~\cite{trugman85} On
the other hand, $\Phi_{4}^2$ is not necessarily an edge eigenstate for
a generic interaction, but the overlap with such a state can be very
close to one. For example, in an $N=6$ system with isotropic
dipole-dipole interaction, the overlap of the corresponding state with
$\Phi_{4}^2$ is 99.85\%. From the form of the wavefunctions, as well
as the linear energy-momentum relation, of the states $\Phi_{2}^2$ and
$\Phi_{4}^{2}$, the latter can be understood as the corresponding
two-boson excitation of the former.  We also point out that these edge
state wavefunctions are not normalized but their normalization can be
straightforwardly worked out (see Appendix~\ref{app:proof}).

\subsection{Ground-state wavefunctions in the anisotropic 
case}

We now turn to the anisotropic IQH ground state in a harmonic trap,
which, for small anisotropy, should be perturbed from the isotropic
state at the edge for energetic reasons. Therefore, the anisotropic
IQH state is expected to be the superposition of the isotropic IQH
ground state and its edge states. Not to be confused with the edge
states for the anisotropic ground state, we may call those for the
isotropic ground state, which have been discussed in the previous
subsection, as {\it isotropic edge states}. The questions in the
following are whether all isotropic edge states contribute, and, if
not, how to describe the subset that contributes in a meaningful way.

First of all, due to the nature of the dipole-dipole interaction, only
isotropic edge states with an angular momentum difference from the
ground state by an integral multiple of 2 can be generated. Therefore,
the leading correction must come from either $\Phi_{2}^{1}$ or
$\Phi_{2}^{2}$ or both. Numerical analysis reveals that the ground
state has vanishingly small overlap with $\Phi_{2}^{1}$, but
substantially large overlap with $\Phi_{2}^{2}$. This is not
surprising, as $\Phi_{2}^{1}$ is a center-of-mass excitation, which is
suppressed by the edge confining potential in the disk geometry. On
the other hand, $\Phi_{2}^{2}$ distorts the otherwise circular droplet
and gains energy from the anisotropic interaction.

\begin{figure}[tbp]
\centering
\includegraphics[width=3.2in]{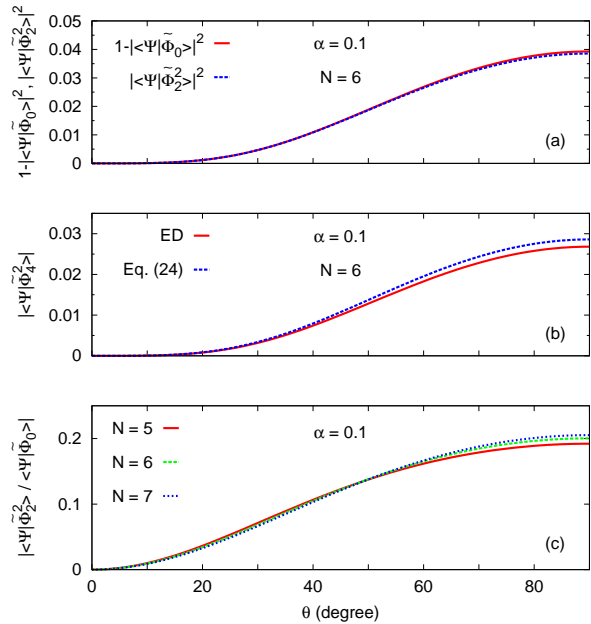}
\caption{(Color online) (a) Comparison of $1 - \vert \langle
  \Psi^{(6)}(\alpha,\theta) \vert \tilde{\Phi}_{0}\rangle \vert^2$ and
  the leading contribution $\vert \langle \Psi^{(6)}(\alpha,\theta)
  \vert \tilde{\Phi}_{2}^{2} \rangle \vert^2$ for a six-particle
  system. (b) Comparison of $\vert \langle \Psi^{(6)}(\alpha,\theta)
  \vert \tilde{\Phi}_{4}^{2} \rangle \vert$ obtained from exact
  diagonalization (ED) and that from the evaluation of
  Eq.~(\ref{eq:connection}), a consequence of the single-mode
  approximation. (c) $\vert \langle \Psi^{(N)}(\alpha,\theta) \vert
  \tilde{\Phi}_{2}^{2} \rangle / \langle \Psi^{(N)}(\alpha,\theta)
  \vert \tilde{\Phi}_0 \rangle \vert$, a measure of the anisotropy
  paramater $\gamma$, for $N = 5$-7. In all three panels we keep
  $\alpha = 0.1$ fixed and vary $\theta$ only. }
\label{overlap}
\end{figure}

We now focus on the regime with $\overline M - M_0 \ll 1$, in which
these edge-state wavefunctions are sufficient for the analysis of
anisotropic IQH states. We choose $\alpha = 0.1$ (the top boundary in
Fig.~\ref{phase}) and calculate the overlaps of the ground state
$\Psi^{(N)}(\alpha,\theta)$ with $N = 6$ obtained from exact
diagonalization and $\tilde{\Phi}_0$, $\tilde{\Phi}_2^2$, and
$\tilde{\Phi}_4^2$, the normalized wavefunctions for $\Phi_{0}$,
$\Phi_{2}^2$, and $\Phi_{4}^{2}$, respectively. In Figure
\ref{overlap}(a) we plot $1 - \vert \langle \Psi^{(6)}(\alpha,\theta)
\vert \tilde{\Phi}_{0}\rangle \vert^2$ and $\vert \langle
\Psi^{(6)}(\alpha,\theta) \vert \tilde{\Phi}_{2}^{2} \rangle \vert^2$.
For $\theta$ between 0 and $90^{\circ}$ the two curves agree very well
and remain small, indicating that the ground state is dominated by
$\Phi_{0}$, and the leading correction is $\Phi_{2}^2$. The next
correction comes from $\Phi_{4}^{2}$. Together, $\Phi_{0}$,
$\Phi_{2}^2$, and $\Phi_{4}^{2}$ exhaust almost the whole $\Psi^{(N)}$
for the strong confinement $\alpha = 0.1$, e.g., up to 99.996\% at
$\theta = 80^{\circ}$. The overlaps of the ground state and other edge
states of the isotropic counterpart are essentially zero.  For
convenience, we also list the values of the significant overlaps at
selected angles $\theta$ in Table~\ref{tbl:overlap}. Therefore, we
conclude that, as $\theta$ increases, the ground state deviates from
$\Phi_0$ to include contributions from $\Phi_{2}^2$ and, to a lesser
extent, $\Phi_{4}^{2}$ (but essentially not the rest).

\begin{table}
\begin{center}
\begin{tabular}{c|cccccccccc}
\hline
$\theta$ & 0$^{\circ}$ & 20$^{\circ}$ & 40$^{\circ}$ & 60$^{\circ}$ & 
80$^{\circ}$ \\\hline\hline
$\vert \langle \Psi^{(6)}(0.1,\theta) \vert \tilde{\Phi}_{0}\rangle \vert$
& 1.0000 & 0.9994 & 0.9945 & 0.9865 & 0.9809 \\\hline
$\vert \langle \Psi^{(6)}(0.1,\theta) \vert \tilde{\Phi}_{2}^2\rangle \vert$
& 0.0000 & 0.0342 & 0.1042 & 0.1628 & 0.1927 \\\hline
$\vert \langle \Psi^{(6)}(0.1,\theta) \vert \tilde{\Phi}_{4}^2\rangle \vert$
& 0.0000 & 0.0008 & 0.0074 &  0.0183 &  0.0258  \\ 
Computed from Eq.~(\ref{eq:connection})
& 0.0000 & 0.0009 & 0.0079 & 0.0195 & 0.0275 \\
\hline
\end{tabular}
\end{center}
\caption{\label{tbl:overlap} Overlaps of the ground state
  $\Psi^{(N)}(\alpha = 0.1,\theta)$ obtained by exact diagonalization
  of an $N = 6$ system and the normalized wavefunctions
  $\tilde{\Phi}_{0}$, $\tilde{\Phi}_{2}^2$, and
  $\tilde{\Phi}_{4}^{2}$. For comparison, we also include the overlap
  with $\tilde{\Phi}_{4}^{2}$ computed from Eq.~(\ref{eq:connection}). }
\end{table}

Not surprisingly, the overlap with the higher-momentum $\Phi_4^2$ is
much smaller than that with $\Phi_2^2$. The question that follows is
whether the former can be related to the latter quantitatively. Given
the general knowledge of trial wavefunctions for correlated systems,
we attemped to fit the numerical wavefunction by
\begin{eqnarray}
\label{eq:fitform}
\Psi_{\gamma} &=& e^{ -\gamma \sum_{i<j}(z_i-z_j)^2} \Phi_{0} \\
&=&\Phi_0 - \gamma \Phi_2^2 + (\gamma^2 / 2) \Phi_4^2 + O(\gamma^3) 
\end{eqnarray}
If this is a good trial wavefunction, we expect the $N$-particle
ground-state wavefunction from exact diagonalization should follow
(see proof in Appendix~\ref{app:proof})
\begin{equation}
\label{eq:connection}
\left \vert \langle \Psi^{(N)} \vert \tilde{\Phi}_{4}^2\rangle \right
\vert = \sqrt{\frac{N^2+1}{2(N^2-1)}} \times \frac {\left \vert
  \langle \Psi^{(N)} \vert \tilde{\Phi}_2^2 \rangle \right \vert^2}
      {\left \vert \langle \Psi^{(N)} \vert \tilde{\Phi}_0 \rangle
        \right \vert}.
\end{equation}
For $N = 6$ the righthand side can be evaluated to be 0.0009, 0.0079,
0.0195, and 0.0275 for $\theta = 20^{\circ}$, 40$^{\circ}$,
60$^{\circ}$, and 80$^{\circ}$, respectively.  The calculated values
are in good agreement with the tabulated ones in
Table~\ref{tbl:overlap}. In Fig.~\ref{overlap}(b) we further compare
$\vert \langle \Psi^{(6)}(\alpha,\theta) \vert \tilde{\Phi}_{4}^{2}
\rangle \vert$ obtained from exact diagonalization and from
Eq.~(\ref{eq:connection}) for the whole range of $\theta$.  The good
agreement is thus consistent with the conjecture that the amplitudes
of $\Phi_0$, $\Phi_2^2$, and $\Phi_4^2$ in $\Psi^{(6)}$ are controlled
by a single parameter $\gamma$. In fact, the simplest way to visualize
the trend of the parameter $\gamma$ as a function of $\theta$ is to
plot $\vert \langle \Psi^{(N)}(\alpha,\theta) \vert
\tilde{\Phi}_{2}^{2} \rangle / \langle \Psi^{(N)}(\alpha,\theta) \vert
\tilde{\Phi}_0 \rangle \vert$, which we show in Fig.~\ref{overlap}(c)
for different system size $N = 5$-7. For practical purposes, it is
worth pointing out that as the two-boson excitation $\Phi_4^2$
amplitude is already small, the overlaps of the ground-state
wavefunction with the three-boson excitation
\begin{equation}
\label{eq:phi62} 
\Phi_{6}^2 = \left [\sum_{i<j}(z_i-z_j)^2
  \right ]^3 \prod_{i<j}(z_i-z_j) e^{-\sum_i|z_i|^2/4},
\end{equation}
and beyond are negligible (unless the edge confinement is too weak).

\begin{figure}[tbp]
\centering
\includegraphics[width=3.2in]{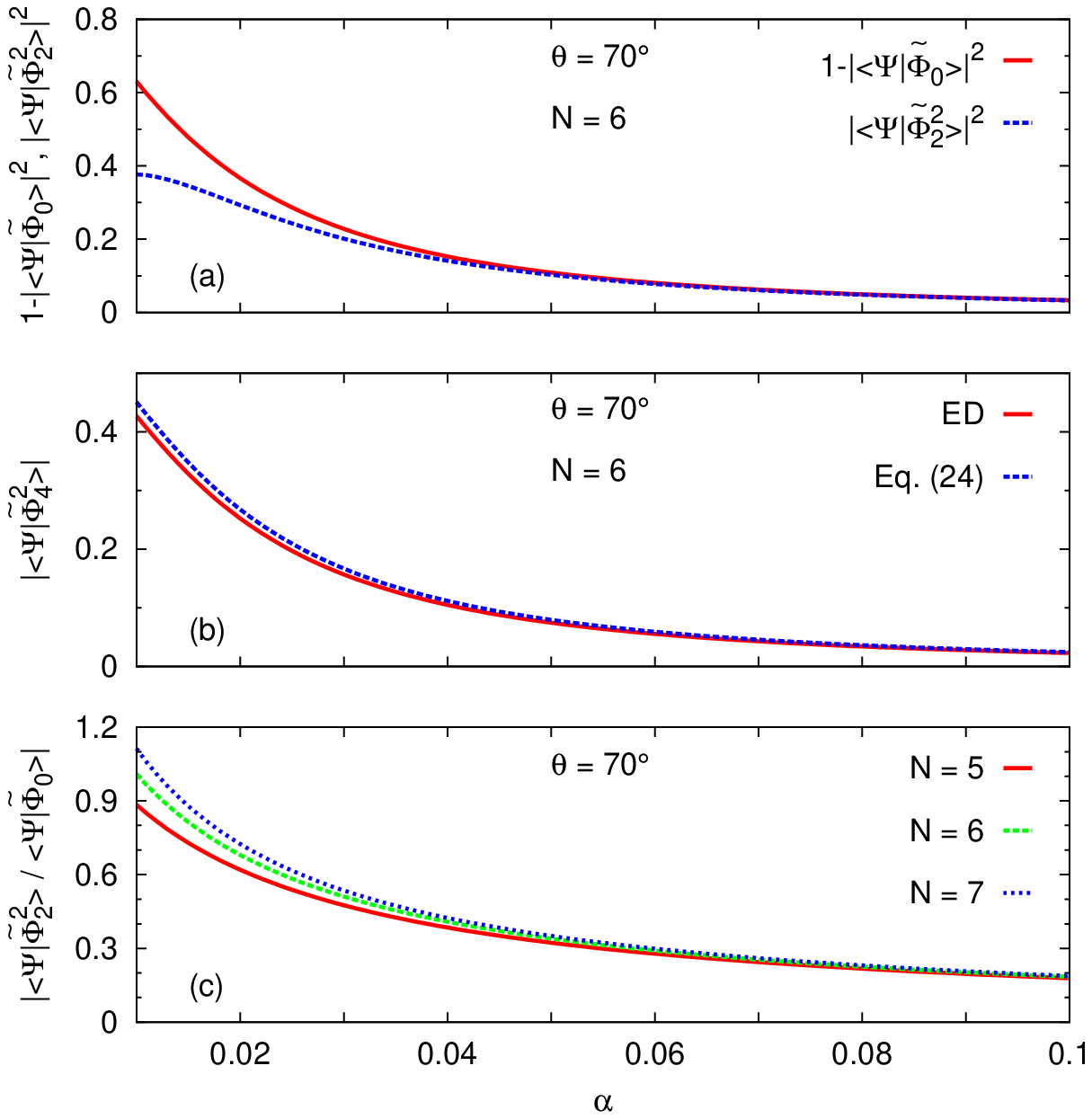}
\caption{(Color online) (a) Comparison of $1 - \vert \langle
  \Psi^{(6)}(\alpha,\theta) \vert \tilde{\Phi}_{0}\rangle \vert^2$ and
  the leading contribution $\vert \langle \Psi^{(6)}(\alpha,\theta)
  \vert \tilde{\Phi}_{2}^{2} \rangle \vert^2$ for a six-particle
  system. (b) Comparison of $\vert \langle \Psi^{(6)}(\alpha,\theta)
  \vert \tilde{\Phi}_{4}^{2} \rangle \vert$ obtained from exact
  diagonalization (ED) and that from the evaluation of
  Eq.~(\ref{eq:connection}), a consequence of the single-mode
  approximation. (c) $\vert \langle \Psi^{(N)}(\alpha,\theta) \vert
  \tilde{\Phi}_{2}^{2} \rangle / \langle \Psi^{(N)}(\alpha,\theta)
  \vert \tilde{\Phi}_0 \rangle \vert$, a measure of the anisotropy
  paramater $\gamma$, for $N = 5$-7. In all three panels we keep
  $\theta = 70^{\circ}$ fixed and vary $\alpha$ between 0.01 (weak
  confinement) and 0.1 (strong confinement). }
\label{overlap2}
\end{figure}

We next turn to the analysis of a sequence of state with increasing
$\alpha$ but fixed $\theta = 70^{\circ}$, in which case we can have
$\overline M - M_0 \sim 1$.  Figure \ref{overlap2} repeats the overlap
analysis for various $\alpha$ at the fixed $\theta$. In
Table~\ref{tbl:overlap2} we list the overlap values for selected
$\alpha$ and, for illustration, the residual weight (up to the second
order) $1 - \vert \langle \Psi^{(N)}(\alpha,\theta) \vert
\tilde{\Phi}_0 \rangle \vert^2 - \vert \langle
\Psi^{(N)}(\alpha,\theta) \vert \tilde{\Phi}_2^2 \rangle \vert^2 -
\vert \langle \Psi^{(N)}(\alpha,\theta) \vert \tilde{\Phi}_4^2 \rangle
\vert^2$. We find that $\Psi^{(N)}(\alpha,\theta)$ is still, to a good
approximation, composed of $\Phi_{0}$, $\Phi_{2}^2$, and
$\Phi_{4}^{2}$. However, when the edge confinement is weak,
higher-order terms may not be negligible and can contribute as much as
7\% for $\alpha = 0.01$ for $N = 6$, in which case the weights of
$\Phi_{0}$ and $\Phi_{2}^2$ are comparable, as indicated in
Fig.~\ref{overlap2}(a). The deviation of $1 - \vert \langle
\Psi^{(6)}(\alpha,\theta) \vert \tilde{\Phi}_{0}\rangle \vert^2$ and
$\vert \langle \Psi^{(6)}(\alpha,\theta) \vert \tilde{\Phi}_{2}^{2}
\rangle \vert^2$ indicates that the higher-order terms are vanishingly
small, as that we have considered earlier.  Nevertheless,
Eq.~(\ref{eq:fitform}) is still a good trial wavefunction, as
indicated by the agreement between $\vert \langle
\Psi^{(N)}(\alpha,\theta) \vert \Phi_{4}^{2} \rangle \vert^2$
calculated from Eq.~(\ref{eq:connection}) and directly from the
overlap calculation, as shown in Fig.~\ref{overlap2}(b). Once again,
the overlap ratio $\vert \langle \Psi^{(N)}(\alpha,\theta) \vert
\tilde{\Phi}_{2}^{2} \rangle / \langle \Psi^{(N)}(\alpha,\theta) \vert
\tilde{\Phi}_0 \rangle \vert$, the measure of $\gamma$, show very weak
size dependence even when the higher-order terms are needed, as
plotted in Fig.~\ref{overlap2}(c).

\begin{table}
\begin{center}
\begin{tabular}{c|cccccccccc}
\hline
$\alpha$ & 0.01 & 0.03 & 0.05 & 0.07 & 0.09 \\\hline \hline
$\vert \langle \Psi^{(6)}(\alpha,70^\circ) \vert \tilde{\Phi}_{0}\rangle \vert$
& 0.6079 & 0.8785 & 0.9440 & 0.9682 & 0.9796 \\
\hline
$\vert \langle \Psi^{(6)}(\alpha,70^\circ) \vert \tilde{\Phi}_{2}^2\rangle \vert$
& 0.6139 & 0.4481 & 0.3210 & 0.2464 & 0.1991 \\
\hline
$\vert \langle \Psi^{(6)}(\alpha,70^\circ) \vert \tilde{\Phi}_{4}^2\rangle \vert$
& 0.4271 & 0.1568 & 0.0745 &  0.0427 &  0.0276  \\ 
Compute from Eq.(\ref{eq:connection})
& 0.4506 & 0.1668 & 0.0794 & 0.0456 & 0.0294 \\
\hline
Residual weight & 0.0712 & 0.0029 & 0.0003 & 0.0001 & 0.0000 \\
\hline
\end{tabular}
\end{center}
\caption{\label{tbl:overlap2} Overlaps of the ground state
  $\Psi^{(N)}(\alpha,\theta = 70^\circ)$ obtained by exact
  diagonalization of an $N = 6$ system and the normalized
  wavefunctions $\tilde{\Phi}_{0}$, $\tilde{\Phi}_{2}^2$, and
  $\tilde{\Phi}_{4}^{2}$. For comparison, we also include the overlap
  with $\tilde{\Phi}_{4}^{2}$ computed from
  Eq.~(\ref{eq:connection}). }
\end{table}

Thus, in this subsection, we have demonstrated that
Eq.~(\ref{eq:fitform}) is a suitable wavefunction to describe the
anisotropic ground state in the IQH regime with dipole-dipole
interaction. The decomposition of the anisotropic wavefunction into a
sum of the isotropic edge-state wavefunctions allows us to further
calculate various properties of the ground state, such as the
expectation value of the total angular momentum (which is no longer an
integer), as well as the edge Green's function. We will not pursue
these straightforward calculations here, but focus on the
understanding of the anisotropy parameter $\gamma$.

\subsection{The single-mode interpretation for the anisotropic 
ground-state wavefunction}

The form of Eq.~(\ref{eq:fitform}) exemplifies the independent-mode
approximation in Eq.~(\ref{eq:jastrow}) in the IQH context. In
particular, since $\Phi_4^2$ is the two-boson excitation of
$\Phi_2^2$, the wavefunction Eq.~(\ref{eq:fitform}) can be regarded as
a single-mode approximation for the anisotropic IQH state. As
discussed in the introduction to the single-mode approximation, we
start from the noninteracting Hamiltonian $H_0 = \alpha L^z$ [the
  first term of Eq.~(\ref{eq:hamiltonian})], which generates the
isotropic IQH state Eq.~(\ref{eq:iqhe}) as its ground state. Our
wavefunction analysis strongly encourage us to assume, in the spirit
of the single-mode approximation, a model quadrupolar interparticle
interaction $V(i, j) = (z_i - z_j)^2$; i.e., $H_1 = \sum_{i<j} V(i,
j)$. The variational wavefunction Eq.~(\ref{eq:fitform}) follows
naturally as we discussed in the introduction. In the more formal
formulation, we can write down a real interparticle interaction
\begin{equation}
V \left (i, j \right ) = (z_i - z_j)^2 + (\bar{z}_i - \bar{z}_j)^2.
\end{equation}
The antiholomorphic contribution drops out after the usual LLL
projection. Not surprisingly, the real form faithfully reproduces the
angular dependence of the anisotropic component of the dipole-dipole
interaction. In the straightforward generalization of the isotropic
pseudopotential to the anisotropic case such a quadrupolar interaction
is the leading correction. It thus remains to be explored how generic
such a model anisotropic wavefunction can be for various interparticle
interactions.

\begin{figure}[tbp]
\centering
\includegraphics[width=3.2in]{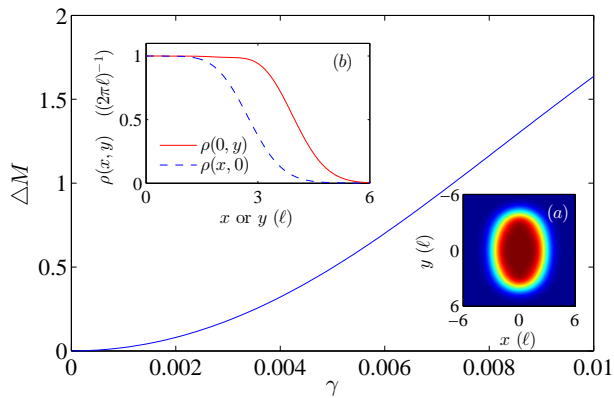}
\caption{(Color online) The increase of the mean angular momentum
  $\triangle M = \langle \Psi_\gamma|L^z|\Psi_\gamma\rangle-M_0$ with
  the anisotropy parameter $\gamma$.  The inset (a) shows the
  elliptical density profile [in units of $1/(2\pi\ell^2)$] on the
  $x$-$y$ plane when $\gamma=0.01$ for an $N=6$ system.  The inset (b)
  shows the density profiles along $x$ and $y$ axes.  }
\label{fig:properties}
\end{figure}

The single-mode approach allows us to write down a close form for the
variational wavefunction, whose general properties can then be
discussed in a quantitative manner. First of all, the rotational
invariance of the wavefunction is broken, which means the total
angular momentum of the state is no longer a good quantum number. In
Fig.~\ref{fig:properties} we plot the continuous change of the
expectation value of the total angular momentum $\triangle M = \langle
\Psi_\gamma|L^z|\Psi_\gamma\rangle-M_0$ as a function of the
anisotropy parameter $\gamma$ for $N = 6$. The plot provides a quick
way to determine $\gamma$ from $\triangle M$ calculated from exact
diagonalization. The increase of $\langle
\Psi_\gamma|L^z|\Psi_\gamma\rangle$ with $\gamma$ comes from the fact
that the interaction anisotropy squeezes the otherwise circular
droplet into an ellipse, as illustrated by the density profile in the
inset (a) of Fig.~\ref{fig:properties} for $\gamma = 0.01$. Away from
the edge, the droplet has the maximum density, as in the isotropic
case. Note that the sign change of $\gamma$ would rotate the ellipse
by $90^{\circ}$. Alternatively, one can show the density profile along
$x$ and $y$ axes as in the inset (b) of
Fig.~\ref{fig:properties}. Regardless of the elliptical shape, one can
still write down the edge states for the anisotropic ground states as
edge density fluctuations. In the same single-mode approximation
framework, we can write down the edge states as the anisotropic ground
state multiplied by symmetrical polynomials. Accordingly, one should
observe the low-lying energy-level countings as 1, 1, 2, 3, 5, $\dots$
for $\triangle M\approx0$, 1, 2, 3, 4, $\dots$, respectively. These
features have already been reported in the exact diagonalization
calculation in Ref.~\onlinecite{qiu11} (Fig. 7). It is worth pointing
out that the variational wavefunctions in the single-mode approach
also allows us to calculate additional properties of much larger
systems by variational Monte Carlo simulation.

\section{Comparison with earlier works} 
\label{sec:discussion}

The consideration of the broken rotational invariance in the quantum
Hall context appeared first in the alternative discussion on the
effect of anisotropic effective mass tensor~\cite{haldaneinbook}.
Very recently the joint effects of both anisotropic mass and
interaction surfaced in the geometric description of quantum Hall
states.~\cite{haldane11} Haldane~\cite{haldane11} pointed out that the
conventional understanding of the Laughlin wavefunctions is not
complete; in his novel geometric description the original Laughlin
wavefunction is simply a member of a family of Laughlin states,
parameterized by a hidden (continuous) geometric degree of freedom.
The family of the Laughlin states, with the geometric factor as a
variational parameter, potentially provides a better description of
the FQH effect in the presence of either anisotropic effective mass or
anisotropic interaction, which are present in real materials.

Haldane~\cite{haldane11} argued that guiding center metric
should be regarded as a variational parameter in the generalized
family of Laughlin trial wavefunctions.  The family of the variational
wavefunction can be constructed by squeezing the LL
orbitals.~\cite{qiu12} In the symmetric gauge one can introduce a
unimodular (or Bogliubov) transformation
\begin{eqnarray}
\left(\begin{array}{c}
z \\
{\bar z} \\
\end{array}\right)
\rightarrow \frac{1}{\sqrt{1-|\lambda|^2}}
\left(\begin{array}{cc}
1 & \lambda^*  \\
\lambda  & 1 \\
\end{array}\right)
\left(\begin{array}{c}
z \\
\bar{z} \\
\end{array}\right).
\end{eqnarray}
of the Landau level (LL) wavefunctions to generate a set of anisotropic
single-particle basis states.~\cite{qiu11}  
The unimodular
transformation approach also appeared in the consideration of Read and
Rezayi~\cite{read11} on the geometric aspects of the uniform
deformation of quantum Hall states in the context of Hall viscocity,
which is the nondissipative transport coefficient that describes the
stress in the adiabatic response to a strain.  The
authors~\cite{read11} pointed out that the extra term in the
exponential corresponds to a quadrupolar harmonic perturbation in the
plasma mapping for the Laughlin state.~\cite{francesco94}

Explicit construction of the family of wavefunctions by the unimodular
transformation~\cite{qiu12} confirms that the isotropic Laughlin
wavefunction is the variational ground state for an isotropic
interparticle interaction. In contrast, the dipole-dipole interaction,
as occurs in rotating dipolar cold atomic systems, leads to an
anisotropic variational ground state.~\cite{qiu12} On a torus geometry
without boundary, one can calibrate the wavefunction anisotropy or the
guiding center metric by exploring static structure factor and pair
correlation function.~\cite{yang12,wang12} The absence of the Jastrow
factor for the isotropic IQH state considered in the present work thus
can be regarded as a special case of the Laughlin state at filling
fraction $\nu = 1$. In the geometric description, the metric-deformed
single-particle basis can be used to set up the stage for a
variational IQH wavefunction~\cite{read11,qiu12}
\begin{equation}
\label{eq:aiqhe}
\Phi_0^{\lambda} (z_1,...,z_N) = e^{-\lambda \sum_i z_i^2 / 4}
\Phi_0(z_1,...,z_N),
\end{equation}
where $\lambda$ parametrizes the metric change or anisotropy.  Note
that the extra term in the exponential, which arises from the
modification of single-particle basis, will be ubiquitous for all FQH
wavefunctions.  Apparently, Eq.~(\ref{eq:aiqhe}) differs from
Eq.~(\ref{eq:fitform}) that we found earlier. However, the single-mode
(or, more generically, independent-mode) approximation here echoes the
geometric description~\cite{haldane11} of quantum Hall states in that
both quest for a (set of) variational parameter(s) that accounts for
the correlation of electrons. It is, therefore, conceivable that for
the IQH effect the two descriptions may be unified, at least for the
IQH state with dipole-dipole interaction. 

Not surprisingly, Eqs.~(\ref{eq:aiqhe}) and (\ref{eq:fitform}) can be
related by noticing
\begin{equation}
\label{eq:decompose}
\sum_i z_i^2 = {1 \over N} \left [\sum_i z_i \right ]^2  + 
{1 \over N} \left [\sum_{i < j} (z_i - z_j)^2 \right ].
\end{equation}
One thus concludes that the relative coordinate part of $\Psi_\gamma$
in Eq.~(\ref{eq:fitform}) is the same (given $\gamma = \lambda / N$)
as that of $\Phi^\lambda_0$ in Eq.~(\ref{eq:aiqhe}), and therefore the
two-body interaction energy for the variational state $\Psi_\gamma$ is
the same as that for the variational state $\Phi^\lambda_0$.  The
difference $e^{-\lambda \left (\sum_i z_i \right )^2/N}$ in the
anisotropic case, when expanded for small $\lambda$, involves
$\Phi_{2}^{1}$ and its descendents, which are not generated by the
dipole-dipole interaction, as we discussed in the wavefunction
analysis.  In fact, these terms vanishes for energetic reasons, as the
expectation value of the one-body part the Hamiltonian
Eq.~(\ref{eq:hamiltonian}) for $\Phi^\lambda_0$ is larger than that
for $\Psi_\gamma$. For generic edge confinement, the presence of
$\exp[-\lambda \left(\sum_i z_i\right)^2/N]$ in $\Phi^\lambda_0$ leads
to larger total angular momentum $L^z$ and larger energy. In the
planer geometry we consider here, one expects that the center-of-mass
coordinate and relative coordinates can be decoupled, and the former
will then be suppressed by the one-body contributions in the
Hamiltonian. Therefore, we conclude that the unimodularly transformed
wavefunction for the IQH case is consistent with the single-mode
approximation of the anisotropic quadrupolar interaction.

\section{Summary}
\label{sec:summary}
In summary, we study the evolution of the ground state wavefunction
for an IQH droplet with anisotropic dipole-dipole interaction. We
perform the numerical anatomy of the ground-state wavefunction and
confirmed that a variational wavefunction can be constructed in the
single-mode approximation in the anisotropic IQH phase as in the
generic treatment for correlated systems. We further argue that the
single-mode approximation for the anisotropic interaction is
consistent with the geometric description for the anisotropic IQH
phase. The present study thus answers, from a different angle, how a
certain perturbation (regarded as additional geometric degree of
freedom) can be accounted for in the trial wavefunction (otherwise
without a variational parameter) for a topological phase. It would be
interesting to see such an understanding can apply to other type of
interactions and other topological phases.

\section*{Acknowledgments}

XW thanks Peter Fulde and Duncan Haldane for the illuminating
discussions that motivated this study. The work was supported by the
973 Program under Project Nos. 2009CB929101 and 2012CB927404, the NSFC
Project Nos. 11174246 and 11274403, and the Discipline Development
Fund Project No. ZDXKFZ201206.

\appendix

\section{Proof of Eq.~(\ref{eq:connection})}
\label{app:proof}

In this appendix we show that if the numerical obtained ground-state
wavefunction $\Psi^{(N)}(\{z_i\})$ for the anisotropic
$N$-particle IQH state can be well described by the variational
wavefunction
\begin{eqnarray}
\label{app:eq:fitform}
\Psi_{\gamma}(\{z_i\}) &=& e^{ -\gamma \sum_{i<j}(z_i-z_j)^2} \Phi_{0} \\
&=&\Phi_0 - \gamma \Phi_2^2 + (\gamma^2 / 2) \Phi_4^2 + O(\gamma^3),  
\end{eqnarray}
where $\gamma$ characterizes the anisotropy, and the unnormalized
isotropic components, as illustrated in the main text, are
\begin{equation}
\label{app:eq:iqhe}
\Phi_0(\{z_i\}) = \left [ \prod_{i < j} (z_i - z_j) \right ]
e^{-\sum_i |z_i|^2/4},
\end{equation}
\begin{eqnarray}
\label{app:eq:phi22}
\Phi_{2}^{2}(\{z_i\})&=& \left [\sum_{i<j}(z_i-z_j)^2
  \right ]\prod_{i<j}(z_i-z_j) e^{-\sum_i|z_i|^2/4} \nonumber \\
&=& 
\left [(N-1)S_2 - (N+1)S_{1,1} \right ] \Phi_0,
\end{eqnarray}
and 
\begin{eqnarray}
\label{app:eq:phi42} 
\Phi_{4}^2(\{z_i\}) &=& \left [\sum_{i<j}(z_i-z_j)^2
  \right ]^2 \prod_{i<j}(z_i-z_j) e^{-\sum_i|z_i|^2/4} \nonumber \\
&=& \left [ (N-1)^2S_{4}-(N-1)(N+3)S_{3,1} \right . \nonumber \\
& & +2(N^2+1)S_{2,2}
-(N-3)(N+1) S_{2,1,1} \nonumber \\
& & \left . +(N+1)^2S_{1,1,1,1} \right ] \Phi_0,
\end{eqnarray}
the overlaps between $\Psi^{(N)}$ and $\Phi_0$, $\Phi_2^2$, and
$\Phi_4^2$ satisfy an equality 
\begin{equation}
\label{app:eq:connection}
\left \vert \langle \Psi^{(N)} \vert \tilde{\Phi}_{4}^2\rangle \right \vert =
\sqrt{\frac{N^2+1}{2(N^2-1)}} \times \frac {\left \vert \langle
  \Psi^{(N)} \vert \tilde{\Phi}_2^2 \rangle \right \vert^2} {\left \vert
  \langle \Psi^{(N)} \vert \tilde{\Phi}_0 \rangle \right \vert},
\end{equation}
where $\tilde{\Phi}_0$, $\tilde{\Phi}_2^2$, and
$\tilde{\Phi}_4^2$ are normalized wavefunctions of $\Phi_0$, $\Phi_2^2$, and
$\Phi_4^2$, respectively.
The proof is basically a bookkeeping of all normalization factors. To
begin with, we approximate the normalized numerical ground state by
\begin{equation}
\Psi^{(N)} = \Psi_{\lambda} / \sqrt{\left \langle \Psi_{\lambda} \vert
  \Psi_{\lambda} \right \rangle } = {\cal N_\lambda} \Psi_{\lambda}.
\end{equation}
with
\begin{eqnarray}
{1 \over {\cal N}_{\lambda}^2} &\equiv& \left \langle \Psi_{\lambda} \vert
  \Psi_{\lambda} \right \rangle \\
&=& 
  \left \langle \Phi_0 \vert \Phi_0 \right \rangle
+ \lambda^2 \left \langle \Phi_2^2 \vert \Phi_2^2 \right \rangle
+ {\lambda^4 \over 4} \left \langle \Phi_4^2 \vert \Phi_4^2 \right \rangle
+ O(\lambda^6). \nonumber
\end{eqnarray}
Given that a Slater determinant ${\mathfrak sl}_{\{ m_i \}}$ in the
FQH context is normalized by
\begin{equation}
{1 \over \sqrt{N!}} \prod_{i = 1}^N {1 \over \sqrt{2 \pi 2^{m_i} m_i!}},
\end{equation}
we obtain 
\begin{eqnarray}
&& \left \langle S_{\{d_i\}} \Phi_0 \vert S_{\{d_i\}} \Phi_0 \right \rangle 
\nonumber \\
&=& N! \prod_{i = 1}^N {(2 \pi) 2^{N-i+d_i} (N-i+d_i)!}
\nonumber \\
&=& 2^{\sum_i d_i} \prod_{i = 1}^N {(N-i+d_i)! \over (N-i)!} 
\left \langle \Phi_0 \vert \Phi_0 \right \rangle.
\end{eqnarray}
Consequently, we find 
\begin{eqnarray}
{\left \langle \Phi_2^2 \vert \Phi_2^2 \right \rangle \over 
\left \langle \Phi_0 \vert \Phi_0 \right \rangle} 
&=& 8 N^2 \left (N^2 -1 \right ), \\
{\left \langle \Phi_4^2 \vert \Phi_4^2 \right \rangle \over 
\left \langle \Phi_0 \vert \Phi_0 \right \rangle} 
&=& 128 N^4 \left (N^4 -1 \right ).  
\end{eqnarray}
With the identification of $\Psi^{(N)} = \Psi_{\gamma} = \Phi_0
- \gamma \Phi_2^2 + (\gamma^2 / 2) \Phi_4^2 + O(\gamma^3)$, we find
\begin{eqnarray}
\langle \Psi^{(N)} \vert \Phi_0\rangle &=& N_{\lambda} \left \langle
\Phi_0 \vert \Phi_0 \right \rangle, \\
\langle \Psi^{(N)} \vert \Phi_2^2 \rangle &=& - \lambda N_{\lambda} \left \langle
\Phi_2^2 \vert \Phi_2^2 \right \rangle \nonumber \\
&=& -8 \lambda N_{\lambda} N^2 (N^2 -1) \left \langle
\Phi_0 \vert \Phi_0 \right \rangle, \\
\langle \Psi^{(N)} \vert \Phi_4^2 \rangle &=& (\lambda^2/2) N_{\lambda} \left \langle
\Phi_4^2 \vert \Phi_4^2 \right \rangle \nonumber \\
&=& 64 \lambda^2 N_{\lambda} N^4 (N^4 -1)  \left \langle
\Phi_0 \vert \Phi_0 \right \rangle. 
\end{eqnarray}
Finally, we reach the equality
\begin{equation}
\left \vert \langle \Psi^{(N)} \vert \tilde{\Phi}_{4}^2\rangle \right \vert =
\sqrt{\frac{N^2+1}{2(N^2-1)}} \times \frac {\left \vert \langle
  \Psi^{(N)} \vert \tilde{\Phi}_2^2 \rangle \right \vert^2} {\left \vert
  \langle \Psi^{(N)} \vert \tilde{\Phi}_0 \rangle \right \vert}.
\end{equation}

\end{document}